\documentclass[sigplan,screen]{acmart}
\usepackage{amsmath,amsfonts}
\usepackage{algorithmic}
\usepackage{array}
\usepackage[caption=false,font=normalsize,labelfont=sf,textfont=sf]{subfig}
\usepackage{textcomp}
\usepackage{stfloats}
\usepackage{url}
\usepackage{verbatim}
\usepackage{graphicx}
\usepackage{multirow}
\usepackage{subcaption}
\usepackage{amsmath,amsfonts, amsthm}
\usepackage{algorithmic}
\usepackage[linesnumbered,ruled,vlined]{algorithm2e}
\usepackage{pifont}
\usepackage{color}
\usepackage{fontawesome}
\usepackage{xcolor,fontawesome}

\usepackage{textcomp}
\usepackage{xcolor}
\usepackage{booktabs}
\usepackage{xspace}
\usepackage{caption}
\usepackage{subcaption}
\usepackage{pgf}
\usepackage{bm}
\usepackage{bbm}
\usepackage{tikz}
\usepackage{comment}
\usepackage{float}
\usepackage{multirow}
\usepackage{enumitem}
\usepackage{multicol}
\usepackage{dsfont}
\usepackage[htt]{hyphenat}

\usepackage{scalerel}
\usepackage{tikz}
\usetikzlibrary{svg.path}
\definecolor{orcidlogocol}{HTML}{A6CE39}
\tikzset{
  orcidlogo/.pic={
    \fill[orcidlogocol] svg{M256,128c0,70.7-57.3,128-128,128C57.3,256,0,198.7,0,128C0,57.3,57.3,0,128,0C198.7,0,256,57.3,256,128z};
    \fill[white] svg{M86.3,186.2H70.9V79.1h15.4v48.4V186.2z}
                 svg{M108.9,79.1h41.6c39.6,0,57,28.3,57,53.6c0,27.5-21.5,53.6-56.8,53.6h-41.8V79.1z M124.3,172.4h24.5c34.9,0,42.9-26.5,42.9-39.7c0-21.5-13.7-39.7-43.7-39.7h-23.7V172.4z}
                 svg{M88.7,56.8c0,5.5-4.5,10.1-10.1,10.1c-5.6,0-10.1-4.6-10.1-10.1c0-5.6,4.5-10.1,10.1-10.1C84.2,46.7,88.7,51.3,88.7,56.8z};
  }
}
\acmYear{2025}
\acmISBN{} 
\acmDOI{} 
\startPage{1}

\setcopyright{none}

\begin{document}

\title[]{Libra: Unleashing GPU Heterogeneity for High-Performance Sparse Matrix Multiplication}

\author{Jinliang Shi, Shigang Li\footnotemark[1], Youxuan Xu, Xueying Wang, Rongtian Fu, Zhi Ma, Tong Wu}

\affiliation{%
  \institution{Beijing University of Posts and Telecommunications}
  \city{Beijing}
  \country{China}
}

\email{shijinliang@bupt.edu.cn, shigangli.cs@gmail.com}

\newcommand*\circled[1]{\tikz[baseline=(char.base)]{\node[shape=circle,fill,inner sep=0.1pt,minimum size=.35cm] (char) {\small{\textcolor{white}{#1}}};}}

\newcommand*\circledw[1]{\tikz[baseline=(char.base)]{\node[shape=circle,draw,fill=white,inner sep=0.1pt,minimum size=.35cm] (char) {\small{\textcolor{black}{#1}}};}}

\newcommand{\cmod}[0]{\;\%\;}  
\renewcommand{\thefootnote}{\fnsymbol{footnote}}

\begin{abstract}
Sparse matrix multiplication operators (i.e., SpMM and SDDMM) are widely used in deep learning and scientific computing. Modern accelerators are commonly equipped with Tensor Core Units (TCUs) and CUDA cores to accelerate sparse operators. The former excels at structured matrix computations, whereas the latter offers greater programming flexibility. However, how to combine these two resources to maximize sparse-operator performance remains unclear. In this work, we first identify the source of performance gains in hybrid computation and systematically analyze their complementary strengths. Motivated by this, we propose Libra, a holistic framework that efficiently leverages heterogeneous computing resources to accelerate both SpMM and SDDMM operators. Specifically, Libra introduces a 2D-aware (locality and utilization) workload distribution method to precisely identify the optimal task mapping, simultaneously leveraging the data reuse capabilities of TCUs and the flexibility of CUDA cores to minimize computational redundancy.
Libra further incorporates hybrid load balancing, occupancy-aware task scheduling, and efficient kernel implementations to maximize execution efficiency. Extensive experiments on H100 and RTX 4090 GPUs demonstrate that Libra surpasses all the 12 up-to-date baselines significantly, e.g., on average $1.77\times$ speedup over FlashSparse, $1.73\times$ over RoDe, and $2.9\times$ over DGL for end-to-end GNN applications. Libra opens up a new perspective for sparse operator acceleration by fully unleashing the power of heterogeneous GPU resources.
\end{abstract}


\maketitle
\footnotetext[1]{ Corresponding author.}

\section{Introduction}
Sparse matrix-dense matrix multiplication (SpMM) and sampled dense-dense matrix multiplication (SDDMM) are two core sparse operators used in various fields, such as deep learning~\cite{heyouni2005matrix,titsias2007infinite,tay2022efficient,wen2018thundersvm,liu2015sparse,scarselli2008graph,kipf2016semi,kumar2022influence} and scientific computing~\cite{lan2014sparse,anzt2015accelerating,blei2003latent}.
For example, in graph neural networks (GNNs), SpMM performs feature aggregation~\cite{bazinska2023cached,xu2018powerful,garcia2017learning}, while SDDMM calculates attention scores between nodes~\cite{thekumparampil2018attention,velivckovic2017graph,huang2005link}.
Currently, two primary technical routes are used to accelerate sparse operators on GPUs. 
The first relies on CUDA cores, which have been extensively explored for sparse computations due to their  programming flexibility~\cite{cuSPARSE, sputnik, RoDe, jiang2020novel}. The second utilizes Tensor Core Units (TCUs), first introduced in NVIDIA’s Volta GPU architecture, which significantly enhance throughput for structured matrix multiplications~\cite{chen2024convstencil,li2022research,khurana2023natural,cusparselt,SparseTC}.
However, TCUs suffer efficiency losses when naively applied to unstructured sparse computations. The strict operand layouts of \textbf{M}atrix-\textbf{M}ultiply-\textbf{A}ccumulate (MMA) operations necessitate zero-padding in registers, resulting in considerable computational redundancy.
To address this issue, TC-GNN~\cite{wang2023tc} introduces SGT, which partitions unstructured sparse matrices into structured $16\times1$ nonzero column vectors, eliminating redundant computations on zero vectors. 
In addition, FlashSparse~\cite{flashsparse} proposes the swap-and-transpose strategy to minimize the granularity of nonzero vectors to $8\times1$, further reducing computational redundancy.
Nevertheless, extremely sparse or highly irregular workloads still incur significant computational redundancy on TCUs. In contrast, CUDA cores can process such cases more efficiently due to their greater programming flexibility, albeit at the cost of lower matrix computation capability.
Consequently, TCUs and CUDA cores each exhibit distinct strengths depending on workload sparsity characteristics.

\begin{figure}[htbp]
  \centering
  \includegraphics[width=\columnwidth]{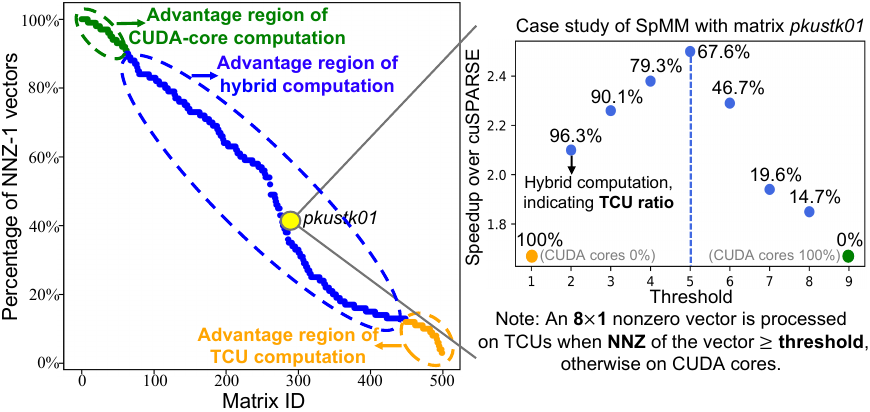}
  \caption{The ratio of NNZ-1 vectors (containing exactly one nonzero element) in all nonzero vectors of a sparse matrix (total 500 matrices from SuiteSparse). The subplot presents a case study showing how Libra’s hybrid SpMM performance varies with the TCU ratio on the matrix \textit{pkustk01}.}
  \label{fig:distribution}
\end{figure}

However, unstructured sparse matrices rarely exhibit a uniform sparsity pattern.
Figure~\ref{fig:distribution} profiles 500 matrices from SuiteSparse~\cite{suitesparse}, measuring the ratio of $8\times1$ column vectors containing exactly one nonzero element (denoted NNZ-1, where NNZ is the number of nonzero elements), which represents maximal computational redundancy on TCUs.
Intuitively, this ratio directly influences the optimal choice of computing resources: high NNZ-1 ratio matrices (green-highlighted region) favor CUDA cores, whereas low NNZ-1 ratio matrices (orange-highlighted region) favor TCUs.
Notably, the wide intermediate region (blue-highlighted) indicates that column vectors of varying sparsity coexist substantially within these matrices, highlighting the potential for performance improvement through hybrid computation.
To validate this insight, we select a representative matrix \textit{pkustk01} from the intermediate region and conduct a case study of SpMM performance using Libra. 
As shown in the subplot of Figure~\ref{fig:distribution}, as the ratio of TCU computation decreases from 100\% (accordingly the ratio of CUDA-core computation increases from 0\%), the execution transitions from TCU-only through hybrid computation to CUDA-core-only.
When the ratio of TCU computation is 67.6\% (i.e., 32.4\% on CUDA cores), the sparse operator achieves the highest performance ($1.4\times$\xspace over the best single-resource implementation), demonstrating the necessity of hybrid computation for achieving high performance in sparse computations.

Recently, several approaches have explored hybrid computation for sparse operators. 
PCGCN~\cite{pcgcn} employs METIS~\cite{metis} to partition dense and sparse subgraphs, distributing each on the optimal computing resource. 
RT-GNN~\cite{yan2025rt} partitions row-wise tiles between TCUs and CUDA cores based on a specific ratio, assigning them within the same thread block for concurrent execution. HC-SpMM~\cite{li2025hc} uses a logistic-regression model to exclusively distribute entire rows of vectors to one resource based on overall sparsity.
Despite these efforts, current methods rarely achieve effective speedups from hybrid computation over single-resource execution. We attribute this limitation primarily to inaccurate workload distribution, suboptimal task scheduling, and inefficient kernel implementations, which stem from two fundamental knowledge gaps of existing works:
\circled{1} It is unclear whether the performance gains of hybrid computation mainly arise from parallel execution, or from fine-grained workload distribution across heterogeneous resources. This ambiguity leads to inefficient utilization of heterogeneous resources to speedup memory-bound SpMM and SDDMM. 
\circled{2} There is no systematic analysis on the complementary strengths of TCUs and CUDA cores. This gap limits the development of the theoretically and empirically optimal workload distribution strategy.
To this end, we propose \textbf{Libra}, a holistic framework that synergistically leverages GPU  heterogeneous resources in a right way to effectively accelerate sparse operators and GNN applications.
We will open source Libra, which is currently under a private repository\footnote{\textcolor{blue}{https://github.com/ParCIS/Libra.git}}.

Our main contributions are: 
\begin{itemize}
\item We are the first to discover the fundamental reason why hybrid computation can bring performance gains and systematically characterize the complementary strengths of TCUs and CUDA cores for accelerating memory-bound SpMM and SDDMM.
\item Building upon theoretical and empirical analysis, we propose a novel 2D-aware (locality and utilization) workload distribution method that precisely identifies the optimal task mapping for SpMM and SDDMM.
\item Libra incorporates hybrid load balancing, occupancy-aware task scheduling, and efficient kernel implementations to maximize sparse computation performance on GPU  heterogeneous resources.
\item Libra significantly outperforms all the 12 up-to-date baselines. Compared to the three best performed baselines, Libra achieves on average speedup of $1.77\times$ over FlashSparse, $1.73\times$ over RoDe, and $2.9\times$\xspace over DGL on end-to-end GNN tasks, with preprocessing overhead as low as 0.4\% of the total runtime.
 \end{itemize}

\section{Background}
\label{sec:TCU}
\subsection{Tensor Core Units}
Tensor Core Units (TCUs)~\cite{tensor} were first proposed in the NVIDIA Volta architecture, and continuously optimized in subsequent architectures such as Ampere~\cite{ampere}, Hopper~\cite{hopper}, and Ada~\cite{ada}.
TCUs specialize in accelerating structured matrix-matrix multiplication~\cite{markidis2018nvidia}, typically performing warp-level MMA operations in the form C = A $\times$ B + C, where operands $A \in \mathbb{R}^{m \times k}$, $B \in \mathbb{R}^{k \times n}$, and accumulator $C \in \mathbb{R}^{m \times n}$.
In an MMA operation, 32 consecutive threads within a warp each load a portion of the operands into their registers and collaboratively compute a shared matrix block.

\subsection{Programming Sparse Operators on TCUs}
\begin{figure}[htbp]
  \centering
  \includegraphics[width=0.9\columnwidth]{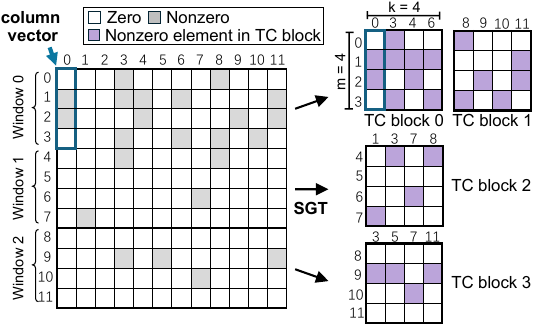}
    \caption{Sparse matrix partition with SGT.}
  \label{fig:b_format}
\end{figure}

Executing sparse operators on TCUs requires identifying nonzero vectors in the sparse matrix to avoid redundant computations on zero vectors~\cite{wang2023tc,dtc,flashsparse}, thereby improving utilization.
Figure~\ref{fig:b_format} exemplifies the case $m=k=4$, where the sparse matrix is partitioned into row windows of size $m$ using SGT~\cite{wang2023tc}. Nonzero elements in each column are then compressed into length-$m$ column vectors, with remaining entries padded with zeros. 
For SpMM, these vectors are condensed by $k=4$, forming a sparse TCU block (\textbf{\textit{TC block}}) of size $m \times k$ as operand A.
For SDDMM, $k$ is replaced by $n$, resulting in an $m \times n$ TC block as accumulator C.

\begin{figure}[htbp]
  \centering
  \includegraphics[width=\columnwidth]{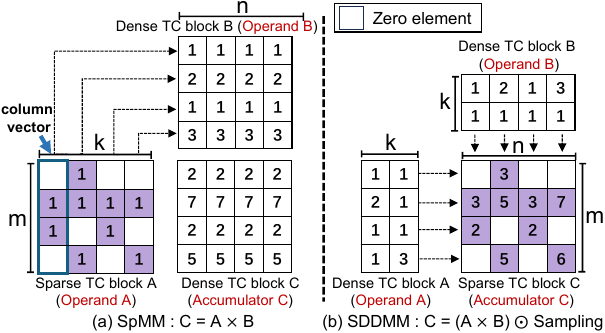}
  \caption{View of MMA operation for SpMM and SDDMM.}
  \label{fig:spmm_sddmm}
\end{figure}

Figure~\ref{fig:spmm_sddmm} illustrates how the partitioned sparse TC block 0 executes the MMA operation in SpMM and SDDMM.
In SpMM, TC block 0 serves as the sparse TC block A (m$\times$k), while the dense TC block B (k$\times$n) is loaded based on the column indices of TC block A. 
The computation results are accumulated in the dense TC block C (m$\times$n).
In SDDMM, TC block 0 serves as the sparse TC block C (m$\times$n).
The dense TC blocks A and B are fetched based on the row and column indices of TC block C.
The computation results are sampled according to the nonzero positions in TC block C.
Due to the irregularity of unstructured sparse data, the number of nonzero elements (i.e., valid computations) varies across different sparse TC blocks, offering opportunities to further improve the efficiency of sparse operators on TCUs.

\section{In-Depth Analysis of Hybrid Computing}
\label{sec:moti}

\subsection{Can Hybrid Computation Bring Superimposed Peak Computational Throughput?}
\label{sec:moti-1}
Although TCUs and CUDA cores can execute concurrently, this does not translate into superimposed (e.g., the sum of the CUDA core peak performance and the TCU peak performance) computational throughput. To empirically validate this conclusion, we conduct controlled experiments with purely computational workloads (i.e., all operands come from registers) on RTX 4090 and H100 GPUs. 
We first establish single-resource performance baselines by exclusively executing either MMA operations on TCUs (TCU-only) or FFMA\footnote{FFMA denotes \textbf{F}P32 \textbf{F}used \textbf{M}ultiply–\textbf{A}dd on CUDA cores.} operations on CUDA cores (CUDA-core-only), capturing each resource’s achievable peak performance.
Next, we establish a Hybrid-concurrent baseline by executing MMA and FFMA simultaneously and vary the workload ratio (defined by the execution time ratio $T_{\text{FFMA}}/T_{\text{MMA}}$) across three cases: CUDA-core-minor workload (ratio=0.1), balanced workload (ratio=1), and CUDA-core-dominant workload (ratio=5).
For further comparison, we also include two hybrid baselines: Hybrid-sequential (executing on one resource at a time) and Hybrid-concurrent-theoretical (theoretical upper bound with perfect performance superposition).

\begin{figure}[htbp]
  \centering
  \includegraphics[width=\columnwidth]{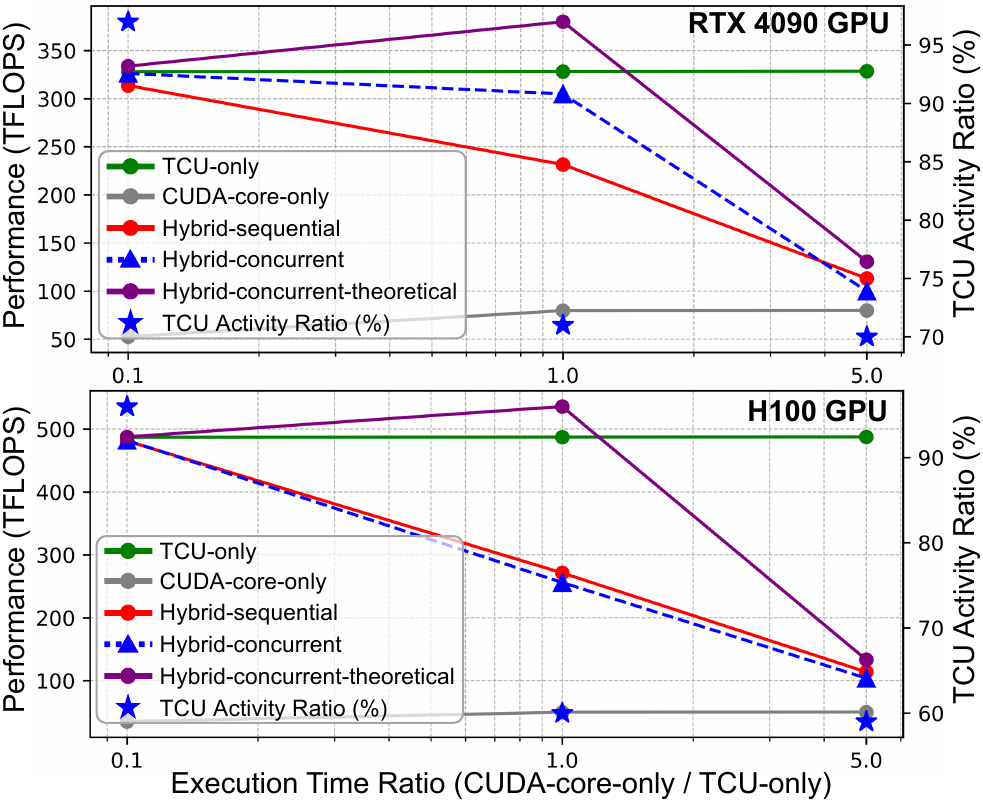}
  \caption{Performance degradation of concurrent execution when increasing CUDA-core workloads. The TCU activity ratio is measured under the Hybrid-concurrent mode.}
  \label{fig:hybrid_profile}
\end{figure}

As shown in Figure~\ref{fig:hybrid_profile}, Hybrid-concurrent execution on the RTX 4090 GPU outperforms Hybrid-sequential under certain workloads, demonstrating that TCUs and CUDA cores indeed execute in parallel.
However, Hybrid-concurrent performance consistently falls short of both the Hybrid-concurrent-theoretical and the TCU-only baselines. 
Profiling with Nsight Systems~\cite{nsight_sys} further reveals that introducing CUDA-core workloads decreases the TCU activity ratio.
\textbf{Consequently, the performance of Hybrid-concurrent is even inferior to TCU-only, since the benefit of concurrent execution fails to compensate for performance degradation caused by lower TCU activity ratio.}
Moreover, extended evaluation on the H100 GPU shows the same conclusion. 
We hypothesize that this performance degradation primarily arises from resource contention between the TCUs and CUDA cores.
Thus, the answer to this section's question is \textbf{No}, i.e., naive concurrency does not deliver any superimposed computational throughput and even lags behind TCU-only execution.

\setlength{\tabcolsep}{1pt}
\begin{table}[]
    \centering
    \caption{SpMM profiling on H100 via Nsight Compute.}
    \begin{tabular}{c cc cc }
    \toprule
    \multirow{2}[2]{*}{Metric} &  \multicolumn{2}{c}{Mip1} &  \multicolumn{2}{c}{Rim}  \\ 
    \cmidrule(lr){2-3} \cmidrule(lr){4-5} 
      & RoDe & FlahSparse & RoDe & FlahSparse \\
    \midrule
    DRAM Load [MB] & 175.1 & 69.77 & 49.65 & 19.71 \\
    Time [$\mu$s] & 265.36 & 156.74 & 63.158 & 53.7 \\
    Mean Bandwidth [GB/s] & 424.69 & 445.13 & 425.88 & 367.03 \\
    Performance [TFLOPS] & 5.02 & 8.50 & 4.11 & 4.83   \\
    \bottomrule
    \end{tabular}
    \label{tab:spmm-ns}
\end{table}

\setlength{\tabcolsep}{1pt}
\begin{table}[]
    \centering
    \caption{SDDMM profiling on H100 via Nsight Compute.}
    \begin{tabular}{c cc cc }
    \toprule
    \multirow{2}[2]{*}{Metric} &  \multicolumn{2}{c}{Mip1} &  \multicolumn{2}{c}{Rim}  \\ 
    \cmidrule(lr){2-3} \cmidrule(lr){4-5} 
      & RoDe & FlahSparse & RoDe & FlahSparse \\
    \midrule
    DRAM Load [MB] & 63.48 & 13.13 & 17.21 & 4.48 \\
    Time [$\mu$s] & 166.05 & 56.1 & 68.03 & 16.4 \\
    Mean Bandwidth [GB/s] & 382.29 & 234.04 & 252.96 & 273.17 \\
    Performance [TFLOPS] & 2.01 & 5.94 & 0.95 & 3.96   \\
    \bottomrule
    \end{tabular}
    \label{tab:sddmm-ns}
\end{table}

\subsection{The Real Strengths of Hybrid Computation}
\label{sec:TCU-reuse}
Next, we will illustrate the real advantages of hybrid computing  for accelerating sparse matrix operators.
Existing works on TCU-based sparse matrix algorithms~\cite{dtc,wang2022qgtc,flashsparse,accspmm,pcgcn,yan2025rt,li2025hc} typically attribute the achieved higher speedup to the higher computing performance of TCUs. However, this is intuitively incorrect since these sparse operators are essentially memory-bound. Different from existing works, our new insight is that \textbf{the architectural advantage of TCUs on data reuse is the key reason to accelerate these memory-bound sparse operators}. Specifically, for data reuse on registers, CUDA-core programming only supports data reuse on thread-local registers, since registers are thread-private. Register exchanging between threads requires explicit shuffle, adding instruction overhead. In contrast, MMA operations on TCUs enable implicit register sharing at the warp level, and the registers from 32 threads in a warp collectively forms coarser grained register blocking (i.e., TC blocks), thus enhancing data reuse on registers. 
To demonstrate our perspective, we select two representative sparse matrices, \textit{Mip1} and \textit{Rim}, from the TCU-advantageous region in Figure~\ref{fig:distribution}. 
Table~\ref{tab:spmm-ns} and Table~\ref{tab:sddmm-ns} quantitatively compare data access costs during SpMM and SDDMM in RoDe~\cite{RoDe} (CUDA-core-based) and FlashSparse (TCU-based), demonstrating that FlashSparse substantially reduces DRAM load due to the improved data reuse on registers. We further analyze this characteristic in Section~\ref{sec:2D}.
The lower mean bandwidth for FlashSparse results from a greater reduction in DRAM load relative to execution time.
However, as sparsity increases towards matrices in the CUDA-core-advantageous region (Figure~\ref{fig:distribution}), most nonzero vectors contain only a single nonzero element.
In this region, the data reuse and utilization of TCUs sharply decline, dropping as low as 12.5\%. 
In such scenarios, \textbf{the advantage of programming flexibility of CUDA cores stands out, which significantly reduces computational redundancy}, highlighting that neither resource alone is universally optimal.
Motivated by these findings, the following sections will present how Libra leverages the complementary strengths of TCUs and CUDA cores to achieve the optimal workload distribution across heterogeneous resources.


\section{Technical Scheme of Libra}
\subsection{Libra Framework Overview}
\begin{figure*}[htbp]
  \centering
  \includegraphics[width=\textwidth]{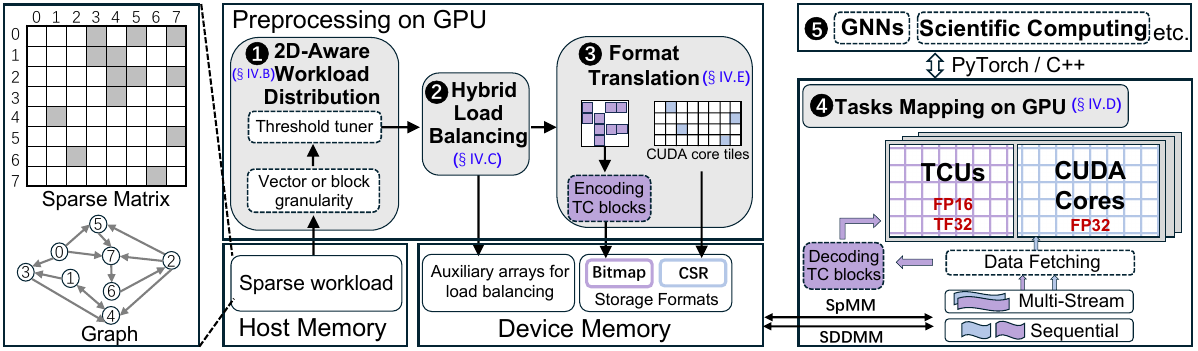}
    \caption{Overview of Libra.}
    \label{fig:overview}
\end{figure*}

Libra is a holistic framework that fully exploits GPU heterogeneity to accelerate sparse matrix multiplication.
As illustrated in Figure~\ref{fig:overview}, Libra consists of several key components, where the first three focus on workload preprocessing on the GPU.
\circled{1} In the 2D-aware workload distribution method, workload is distributed between TCUs and CUDA cores based on two key dimensions: data reuse and practical TCU utilization. 
The former determines distribution granularity, while the latter is guided by a threshold tuner.
\circled{2} Libra employs a hybrid load-balancing strategy to evenly distribute workloads across thread blocks.
\circled{3} The format translation is GPU-accelerated, with the TCU portion encoded in Bitmap format and the CUDA-core portion in CSR format.
For a given matrix, preprocessing is performed only once, and the distribution information can be reused across subsequent iterations.
The next two components focus on the kernel runtime on the GPU.
\circled{4} Libra efficiently maps distributed tasks onto heterogeneous resources through occupancy-aware task scheduling and efficient low-level kernel implementations.
\circled{5} We use pybind11~\cite{pybind11} to encapsulate the highly optimized CUDA kernels, enabling PyTorch to invoke these kernels for accelerating end-to-end GNN tasks.

\subsection{2D-Aware Workload Distribution}
\label{sec:2D}
\begin{figure}[htbp]
  \centering
  \includegraphics[width=\columnwidth]{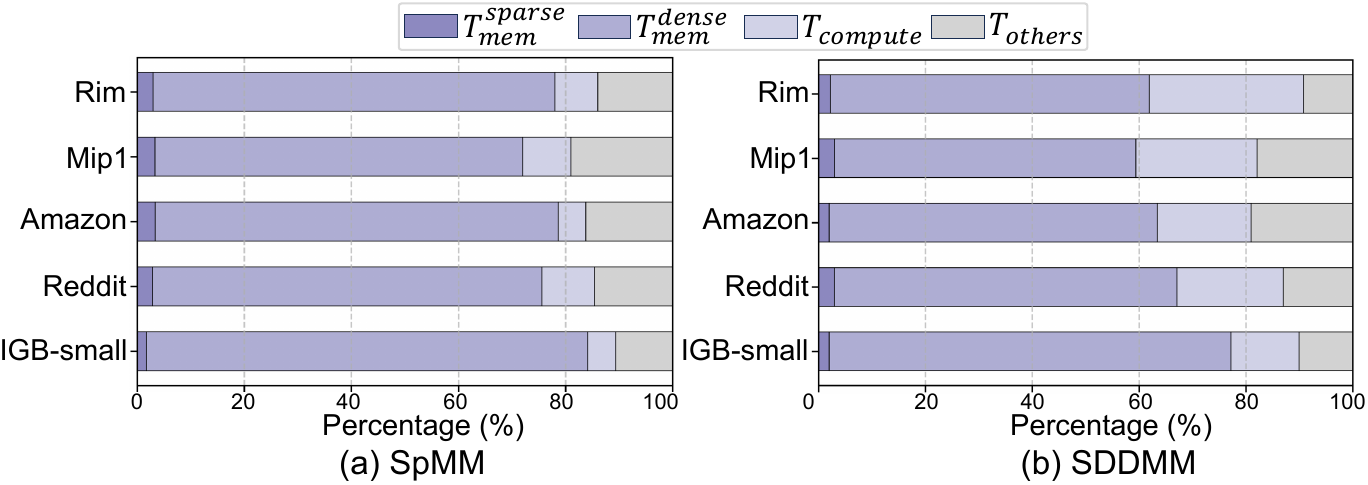}
  \caption{Breakdown of execution time for SpMM and SDDMM on TCUs across different datasets.}
  \label{fig:runtime_breakdown}
\end{figure}

We first break down the execution time of SpMM and SDDMM on TCUs into four primary components to locate performance bottlenecks:
\begin{equation}
T_{\text{SpMM/SDDMM}} = T_{\text{mem}}^{\text{sparse}} + T_{\text{mem}}^{\text{dense}} + T_{\text{compute}} + T_{\text{others}}
\end{equation}
where $T_{\text{mem}}^{\text{sparse}}$ and $T_{\text{mem}}^{\text{dense}}$ denote the access times for sparse and dense matrices, respectively;
$T_{\text{compute}}$ denotes the execution time of MMA operations;
$T_{\text{others}}$ includes essential overheads, such as synchronization and result write-back.
As shown in Figure~\ref{fig:runtime_breakdown}, $T_{\text{mem}}^{\text{sparse}}$ minimally impacts the overall runtime, whereas $T_{\text{mem}}^{\text{dense}}$ clearly dominates.
This is because, in both SpMM and SDDMM, sparse-matrix access volume primarily scales with the number of nonzeros, whereas dense matrices incur significantly greater memory traffic, scaling with both the number of nonzeros and the dense-matrix dimensions. Furthermore, dense-matrix accesses typically exhibit irregular patterns, as non-contiguous dense rows are fetched based on sparse indices.
\textbf{Thus, dense matrix access is the primary source of the data access bottleneck in SpMM and SDDMM}. To better understand and mitigate this bottleneck, we analyze the distinct data reuse patterns exhibited by TCUs and CUDA cores.
In addition, the computation phase ($T_{\text{compute}}$), primarily governed by actual TCU efficiency, also affects the total runtime.
Thus, our workload distribution method focuses on two key dimensions across different computing resources and sparse operators: \textbf{Data reuse}, targeting the dominant $T_{\text{mem}}^{\text{dense}}$, and \textbf{Practical TCU utilization}, determining $T_{\text{compute}}$.

\subsubsection{\textbf{Data Reuse.}} 
As discussed in Section~\ref{sec:TCU-reuse}, TCUs exhibit a distinctive architectural advantage in data reuse. Specifically, during MMA operations, each thread only supplies a portion of the operand fragments stored in its local registers, which are then shared implicitly to other threads within the same warp~\cite{a100white}. This eliminates explicit data-movement instructions and greatly mitigates the data access bottleneck associated with dense operands in SpMM and SDDMM.
In contrast, CUDA cores do not support such hardware-implicit register sharing and instead rely on explicit shuffle instructions (e.g., \textit{\_\_shfl\_sync}) for intra-warp register data exchange, which introduces additional instruction overhead. 
Existing CUDA-core-based implementations~\cite{sputnik,RoDe} typically load dense operands directly into registers from global memory, depending on caches for implicit reuse (shared-memory reuse is generally impractical as dense operands exceed shared-memory capacity~\cite{dtc}). Cache-based reuse can alleviate global memory traffic but still requires explicit transfers into registers, adding both instruction overhead and potential latency from memory-hierarchy contention.
Therefore, to facilitate analysis, we define a unified metric, \textbf{\textit{data access cost}}, which refers to the cost of fetching operands from the memory hierarchy when they are not already resident in the consumer thread’s registers, without distinguishing the data source.

For SpMM, we define the data access cost ratio between CUDA cores and TCUs for accessing dense TC block B (shown in Figure~\ref{fig:spmm_sddmm}(a)) as:
\begin{equation}
R_{spmm} = \frac{\text{NNZ} \times n}{k \times n} = \frac{\text{NNZ}}{k}
\label{eq:equation_spmm}
\end{equation}
where NNZ means the \textbf{N}umber of \textbf{N}on-\textbf{Z}ero elements in the sparse TC block A; $m$, $n$ and $k$ are the dimensions of MMA operands. For CUDA cores, the data access cost is $\text{NNZ}\times n$.
This is because each nonzero element, processed individually, must be multiplied with an entire row of the dense TC block B.
In contrast, for TCUs, the cost is $k \times n$.
This is because each row of the dense TC block B is loaded into registers once and then reused across the nonzero elements within the same column vector. Consequently, $R_{spmm}$ depends on NNZ and $k$. When $\text{NNZ} > k$, TCUs can reduce the data access cost by a factor of $\frac{\text{NNZ}}{k}$ because of data reuse. Furthermore, we use $\rho$ to represent the density of the sparse TC block A ($m \times k$). Substituting $\text{NNZ} = mk\rho$ into Equation~(\ref{eq:equation_spmm}):
\begin{equation}
R_{spmm} = \frac{m k \rho}{k} = m\rho
\label{eq:equation_spmm2}
\end{equation}
where $m\rho$ is the average number of nonzeros across all nonzero column vectors in TC block A. Intuitively, a vector with higher density benefits more from data reuse when processed on TCUs. Therefore, \textbf{for SpMM}, the sparse workload is distributed between TCUs and CUDA cores at the granularity of \textbf{nonzero column vectors} (i.e., $m\times 1$).

However, for SDDMM, the scenario differs significantly, as both input TC blocks A and B are dense, as shown in Figure~\ref{fig:spmm_sddmm}(b). This necessitates a distinct analysis of data reuse. Specifically, the ratio of data access costs between CUDA cores and TCUs can be expressed as follows:
\begin{equation}
R_{sddmm} = \frac{2 \times \text{NNZ} \times k}{m \times k + n \times k}
= \frac{2 \times \text{NNZ}}{m + n}
\label{eq:equation_sddmm}
\end{equation}
where NNZ means the Number of Nonzero elements in the sparse TC block C. For CUDA cores, the data access cost is $2\times\text{NNZ}\times k$. This is because calculating each nonzero element in TC block C needs to access both a row from TC block A and a column from TC block B. However, when using TCUs, TC blocks A and B are only loaded once, and reused in the MMA operation. Ultimately, $R_{sddmm}$ depends on NNZ, $m$, and $n$. When $\text{NNZ} > \frac{m+n}{2}$, TCUs can reduce the data access cost by a factor of $R_{sddmm}$ because of data reuse. Unlike SpMM, Equation~(\ref{eq:equation_sddmm}) cannot be further simplified. Here, $m$ and $n$ are fixed dimensions of MMA operands. Thus, the sparse TC block C containing more nonzero elements benefits from better data reuse when processed on TCUs. Therefore, \textbf{for SDDMM}, the sparse workload is distributed between TCUs and CUDA cores at the granularity of \textbf{TC blocks} (i.e., $m\times n$).

In SpMM, since each $m \times 1$ nonzero column vector contains at least one nonzero element, we have $m\rho \geq 1$. 
In SDDMM, each $m \times n$ sparse TC block C is composed of multiple $m \times 1$ nonzero column vectors, containing at least $n$ nonzero elements. Therefore, we have $\text{NNZ} \geq n$, except possibly for incomplete TC blocks at the end of row windows. 
Additionally, since we employ a fine-grained workload-to-MMA-operand mapping~\cite{flashsparse}, the operand dimensions of TC block C typically satisfy $m < n$.
Substituting these constraints into Equations (\ref{eq:equation_spmm2}) and (\ref{eq:equation_sddmm}):
\begin{equation}
R_{spmm} \geq 1 \; ; \qquad
R_{sddmm} \geq \frac{2 \times n}{m + n} > 1
\end{equation}
Therefore, the condition for TCUs to achieve lower data access costs compared to CUDA cores can be easily satisfied.

\subsubsection{\textbf{Practical TCU Utilization.}}
\label{sec:practice}
However, satisfying the data reuse condition alone ($R > 1$) is insufficient for choosing TCUs over CUDA cores. 
This is because TCUs enforce a strict operand layout for MMA operations, necessitating zero-padding. Thus, if the fraction of nonzero elements is excessively low, TCUs perform redundant zero computations, reducing utilization. Consequently, we introduce a practical TCU utilization metric $\eta$ for SpMM and SDDMM:
\begin{equation}
\begin{cases}
\eta_{\text{spmm}} = \dfrac{ \text{NNZ}_{\text{vector}} }{ m }, & 0 < \eta_{\text{spmm}} \leq 1 \\[10pt]
\eta_{\text{sddmm}} = \dfrac{ \text{NNZ}_{\text{block}} }{ m \times n }, & 0 < \eta_{\text{sddmm}} \leq 1
\end{cases}
\label{eq:conditions1}
\end{equation}
where $\text{NNZ}_{\text{vector}}$ and $\text{NNZ}_{\text{block}}$ denote the number of nonzero elements in a column vector (SpMM) and a TC block (SDDMM).
A higher $\eta$ (approaching 1) indicates a greater proportion of effective arithmetic operations executed by the TCU. 
Therefore, we introduce a utilization threshold $\eta_{thr}$ and assign a workload to the TCU only if it satisfies $\eta \geq \eta_{thr}$,
otherwise, it is distributed to CUDA cores.
Since the practical computational throughput on TCUs can be estimated as the theoretical peak performance (determined by hardware) scaled by $\eta$, we hypothesize that the optimal threshold $\eta_{thr}$ is predominantly influenced by hardware architecture rather than individual matrices. This hypothesis aligns with our empirical results in Section~\ref{sec:threshold_eva}.
Compared with existing hybrid approaches (e.g., PCGCN relying on METIS tuning, HC-SpMM using a logistic-regression model), Libra drastically reduces the search space, making hybrid execution practical.

\begin{figure}[htbp]
  \centering
  \includegraphics[width=\columnwidth]{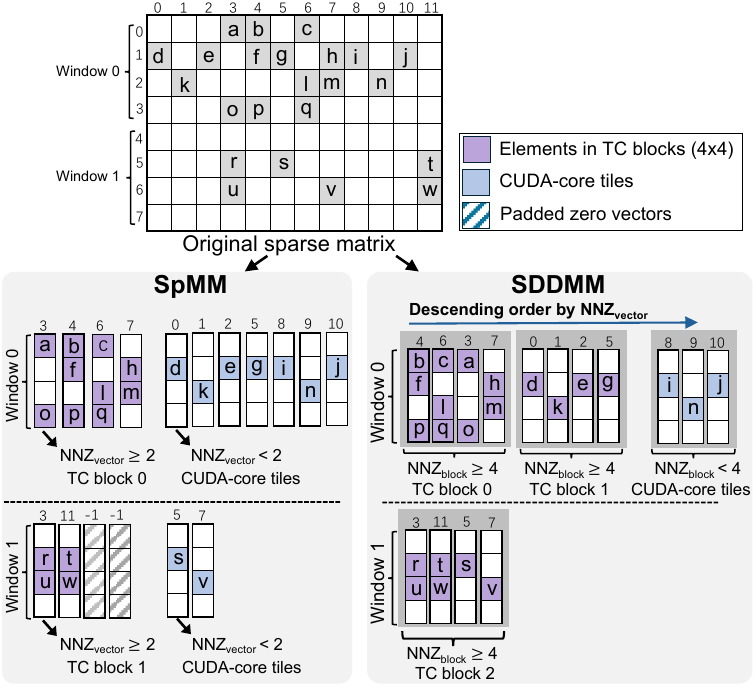}
  \caption{Illustration of 2D-Aware Workload Distribution. Example thresholds: $\eta_{thr}=0.5$ (i.e., $\text{NNZ}_{\text{vector}}=2$) for SpMM, and $\eta_{thr}=0.25$ (i.e., $\text{NNZ}_{\text{block}}=4$) for SDDMM.
}
  \label{fig:partition}
\end{figure}

Figure~\ref{fig:partition} exemplifies the 2D-aware workload distribution for SpMM and SDDMM using a 4$\times$4 TC block for clarity.
For SpMM, we set $\eta_{thr}=0.5$. We first count the NNZ in each column vector within every window. If NNZ$\geq$2 (i.e., $\eta \geq 0.5$), the vector is assigned to TCUs (purple-highlighted); otherwise, it is assigned to CUDA cores (blue-highlighted). TCU-assigned vectors are typically condensed into 4$\times$4 TC blocks to match the MMA operand layout, with zero vectors as padding. Padded zero vectors can be replaced by CUDA-core-assigned vectors.
For SDDMM, we set $\eta_{thr} = 0.25$. \textbf{We first sort column vectors within each window in descending $\text{NNZ}_{\text{vector}}$}, condensing the densest vectors into 4$\times$4 TC blocks. If NNZ in a TC block $\geq$ 4 (i.e., $\eta \geq 0.25$), the TC block is assigned to TCUs; otherwise, it is assigned to CUDA cores.
In practice, with the swap-and-transpose strategy~\cite{flashsparse}, Libra adopts a vector granularity of 8$\times$1 for SpMM and a TC block granularity of 8$\times$16 for SDDMM.
Overall, our strategy combines theory and practice, enabling precise workload distribution on GPU  heterogeneous resources.

\subsection{Hybrid Load Balancing}
\label{sec:load}
Following workload distribution, the next step is to evenly assign these workloads across thread blocks in parallel systems. However, load balancing has received limited attention in hybrid computation methods~\cite{pcgcn,yan2025rt,li2025hc}. Existing approaches typically balance workloads separately for either TCUs or CUDA cores, whereas Libra explicitly balances both simultaneously.
After 2D-aware workload distribution, certain windows might contain excessive TC blocks or overly long CUDA-core tiles, requiring further decomposition. However, each segment from window decomposition in SpMM introduces \texttt{atomicAdd} instructions. We observe that decomposing windows with few workloads or enforcing uniform partitioning via cross-window decomposition~\cite{dtc} often yields insufficient benefits to justify atomic overhead.
To balance this trade-off, Libra employs two threshold group sizes, $T_s$ and $C_s$, and triggers decomposition only when the workload exceeds these thresholds.

\begin{figure}[htbp]
  \centering
  \includegraphics[width=\columnwidth]{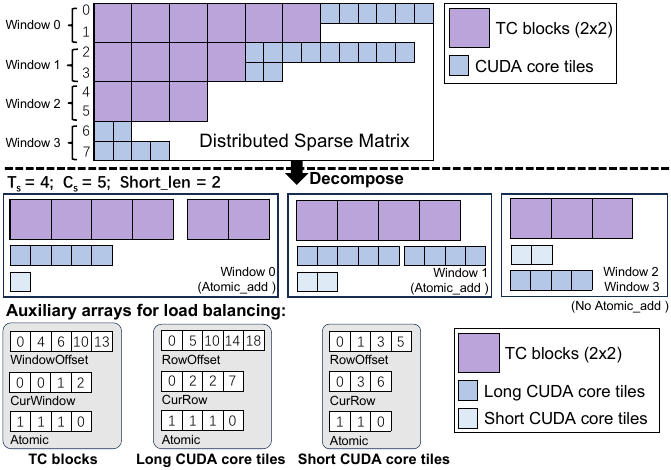}
  \caption{Illustration of hybrid load balancing. $T_s$ and $C_s$ are the threshold group sizes of TCU and CUDA cores.}
  \label{fig:windowsplit}
\end{figure}

Figure~\ref{fig:windowsplit} exemplifies the window decomposition with TC blocks (e.g., 2$\times$2) and CUDA core tiles based on the specified decomposition criteria.
For TC blocks, we decompose each window into TC block groups, each consisting of $Ts$ TC blocks (we use $Ts=4$ as an example). 
For CUDA core tiles, we employ the long and short tile division method~\cite{RoDe}. Specifically, we introduce a threshold parameter $Short\_len$ to classify CUDA core tiles as short and long CUDA core tiles (with $Short\_len=2$ as an example). Long CUDA core tiles are further partitioned into CUDA core tile groups, each containing $Cs$ CUDA core tiles (we use $Cs=5$ as an example).
We summarize three cases of window decomposition in Libra: 
\textbf{In window 0}, since both the TC blocks and long CUDA core tiles need to be decomposed, all segments in this window require atomic operations; 
\textbf{In window 1}, since the CUDA core tiles need to be decomposed, the TC blocks also require atomic operations. Similarly, once the TC blocks need to be decomposed, the CUDA core tiles also require atomic operations; 
\textbf{In windows 2 and 3}, since these windows contain a single type of workload and do not meet the decomposition criteria, atomic operations are not required.

Furthermore, we introduce three auxiliary arrays to record the decomposition information.
\circledw{1} \textit{WindowOffset} and \textit{RowOffset} record the number of TC blocks and non-zero elements in each segment, respectively. 
\circledw{2} \textit{CurWindow} and \textit{CurRow} track the original window and row indices for each segment before decomposition.
\circledw{3} \textit{Atomic} indicates whether each segment requires atomic operations.
Despite the atomic-operation overhead, properly chosen thresholds ensure that load-balancing benefits outweigh atomic costs. 
We empirically set $T_s=16$, $C_s=32$, and $Short\_len=3$, which provide robust load balancing for the majority of matrices (validated in Table~\ref{tab:ab_format}).

\subsection{Task Mapping in Hybrid Computation}
\label{sec:map}
After offline preprocessing, the next challenge is efficient task mapping to heterogeneous GPU resources at runtime.
\subsubsection{Occupancy-Aware Task Scheduling.}
Previous hybrid approaches~\cite{yan2025rt, li2025hc} schedule TCU and CUDA-core tasks directly within the same CUDA kernel or thread block, aiming for concurrent execution over closely spaced instruction cycles.
However, as analyzed in Section~\ref{sec:moti-1}, such concurrency often fails to deliver performance superposition. More importantly, TCUs and CUDA cores exhibit distinct execution characteristics: the former is optimized for block-level MMA operations, whereas the latter specializes in handling fine-grained scalar and vector instructions. A unified kernel-launch configuration (e.g., gridDim and blockDim) forces both resources into a common execution configuration, thereby limiting achievable efficiency in hybrid computation.
To overcome these limitations, we decouple the two execution paths and map them onto separate CUDA streams. This multi-stream scheduling not only enables independent kernel optimizations but also enhances SM utilization. 
Nevertheless, in SpMM, multi-stream scheduling introduces inter-path atomic overhead, arising when workloads from TCUs and CUDA cores coexist within the same window (e.g., window 0 in Figure~\ref{fig:windowsplit}). For large-scale sparse matrices, at least one execution path usually saturates all SMs. In such cases, sequential scheduling is preferable as it avoids inter-path atomic overhead.
Building on these trade-offs, we propose an occupancy ratio metric $O$ that guides the selection between multi-stream and sequential scheduling modes:
\begin{equation}
O = \frac{B_{\text{launch}}}{G_{\max}}, 
\quad G_{\max} = N_{\text{SM}} \times B_{\max}^{\text{SM}}
\end{equation}
where 
$N_{\text{SM}}$ represents the total number of SMs on the GPU;
$B_{\max}^{\text{SM}}$ is computed by the CUDA occupancy API as the maximum thread blocks per SM based on kernel resource usage;
$B_{\text{launch}}$ denotes the total number of thread blocks launched per execution path, computed as $ \prod_{i\in\{x,y,z\}}\texttt{gridDim.}i$.

To determine scheduling mode, we first calculate occupancy ratios $O$ for each execution path.
Then, using empirically calibrated occupancy thresholds ($O_{\text{thr}}^{\text{TCU}}$ and $O_{\text{thr}}^{\text{CUDA}}$) obtained through profiling (e.g., $O_{\text{thr}}^{\text{TCU}}\approx 3.91$, $O_{\text{thr}}^{\text{CUDA}}\approx 38.27$ for H100 GPU), we normalize these occupancy ratios:
\begin{equation}
  \hat O_{\text{TCU}}=\frac{O_{\text{TCU}}}{O_{\text{thr}}^{\text{TCU}}},\quad
  \hat O_{\text{CUDA}}=\frac{O_{\text{CUDA}}}{O_{\text{thr}}^{\text{CUDA}}}
\end{equation}
\textbf{Scheduling decision:}  
If $\max(\hat O_{\text{TCU}},\hat O_{\text{CUDA}})<1$, indicating both execution paths are under-saturated, we select \emph{multi-stream}; otherwise, we select \emph{sequential} scheduling.
The threshold settings are typically architecture-dependent. For a new GPU architecture, the thresholds only need to be determined once through preprocessing. 

\subsubsection{Kernel Implementation.}
To achieve optimal hybrid computation performance, we provide customized SpMM and SDDMM kernels in Libra. For TCUs, we leverage low-level PTX MMA operations for various operand shapes, specifically \textit{m16n8k8} and \textit{m16n8k16}. 
To enable fine-grained column-vector granularity and enhance MMA utilization, we employ the swap-and-transpose strategy~\cite{flashsparse}. 
We execute MMA in TF32/FP16 with FP32 accumulation, preserving output consistency with CUDA-core kernel (FP32).
To further improve efficiency, we leverage the Bitmap data format~\cite{fan2025spinfer,accspmm,xia2025voltrix} and propose two novel optimizations over previous methods.
\circledw{1} We observe that elements within the same TC block are encoded into a single 32/64-bit Bitmap element. Consequently, threads within the same warp simultaneously access this common element, achieving high temporal locality. Leveraging this property, our kernel implementation directly loads compressed bit-encoded elements from global memory into registers and performs decoding in registers, completely bypassing shared memory and eliminating synchronization overhead within thread blocks.
\circledw{2} Previous approaches have primarily leveraged the Bitmap format to accelerate SpMM, leaving the potential of Bitmap in SDDMM unexplored. Compared to SpMM, SDDMM additionally requires computing intra-block offsets of nonzero elements during write-back.
To address this, we extend the Bitmap approach to SDDMM, utilizing the direct mapping between bits and thread IDs. By invoking the CUDA built-in function \texttt{popc()}, intra-block offsets are efficiently computed without iterative traversal.
For CUDA-core execution, we employ coalesced and vectorized memory operations (float4, half2) to optimize data access efficiency.
Overall, these execution-path-specific kernel optimizations ensure the full unleashing of heterogeneous resources.

\subsection{GPU-accelerated Preprocessing Algorithm}
The preprocessing overhead in Libra primarily involves 2D-aware workload distribution strategy, hybrid load balancing, and the associated data format translation.
We summarize the preprocessing algorithm into three stages.
\circledw{1} Each CUDA thread records the window and column indices of each non-zero element from the sparse matrix into memory arrays, preparing for hybrid workload distribution.  
\circledw{2} The \textit{generate\_Distribution\_Information} function is invoked to efficiently distribute the hybrid workload and achieve balanced mapping across thread blocks.
\circledw{3} Based on the distribution information and target storage formats (i.e., Bitmap and CSR formats), global memory is allocated for the result arrays, and dynamic shared memory is utilized to populate the arrays with the distributed information.

The detailed implementation of Stage\circledw{2} is shown in Algorithm~\ref{algo:preprocess}.
First, each CUDA thread is mapped to handle the distribution information for the TCU portion of each window. Next, the column indices of the column vectors within the TCU portion are updated. Then, each CUDA thread is mapped to the remaining non-zero elements in each row within the window, ultimately generating the distribution information for the CUDA core tiles.
Overall, although the preprocessing is achieved through complex and multiple kernel launches, the powerful parallelism of CUDA threads and the negligible overhead of kernel launches enable us to achieve significant acceleration compared to CPU-based strategies, especially when handling extremely large sparse matrices.

\begin{algorithm}
  \caption{generate\_Distribution\_Information}\label{algo:preprocess}
  \small
  \DontPrintSemicolon
  \KwData{Sparse matrix and parameters $\eta_{thr}$, $Short\_len$, $Ts$, $Cs$}
  \KwResult{The distributed information of TCUs and CUDA core tiles}
  
  \textcolor{blue}{//\textbf{Step1:} Call CUDA function $distribute\_TCU\_portion()$;}\;
  \textbf{Parallel for each CUDA thread:} \;
  \ForEach{thread $i$ in windows}{
    \textbf{Let} $v_i$ \text{be the column vector assigned to thread } $i$\;
    \If{Number of elements in $v_i$ $>$ $\eta_{thr}$}{
        \textbf{Distribute elements of } $v_i$ \text{ as TCUs portion}\;
        \textbf{Decompose the TCUs portion based on $Ts$}\;
        \textbf{Record element information of TCUs portion}\;
    }
  }
  
  \textcolor{blue}{//\textbf{Step2:} Update column indices of column vectors in TCUs portion;}\;

  \textcolor{blue}{//\textbf{Step3:} Call CUDA function $distribute\_CUDA\_core\_tiles()$;}\;
  \textbf{Parallel for each CUDA thread:} \;
  \ForEach{thread $i$ in windows with elements outside TCUs portion}{
    \ForEach{row $r_j$ in cur\_window}{
      \textbf{Count the number of elements in row $r_j$}\;
      \If{The number of elements in $r_j$ is less than $Short\_len$}{
          \textbf{Distribute row $r_j$ as a short row}\;
          \textbf{Record element information of CUDA\_short\_tiles}\;
      }
      \Else{
          \textbf{Decompose the CUDA\_long\_tiles based on $Cs$}\;
          \textbf{Record element information of CUDA\_long\_tiles}\;
      }
    }
  }
\end{algorithm}

\section{Evaluation}

\subsection{Experimental Setup}
We conducted our experiments on two NVIDIA GPU architectures: NVIDIA H100 PCIe (Hopper with 80 GB) and GeForce RTX 4090 (Ada Lovelace with 24 GB). 

\textbf{Baselines: }We compare Libra with state-of-the-art works:
\circledw{1} Works on TCUs: \textbf{TC-GNN}~\cite{wang2023tc}, \textbf{DTC-SpMM}~\cite{dtc}, and \textbf{FlashSparse}~\cite{flashsparse};  
\circledw{2} Works on CUDA cores: \textbf{cuSPARSE}~\cite{cuSPARSE}, \textbf{Sputnik}~\cite{sputnik}, and \textbf{RoDe}~\cite{RoDe};
\circledw{3} Works on heterogeneous resources: \textbf{HC-SpMM}~\cite{li2025hc}, \textbf{RT-GNN}~\cite{yan2025rt}, and \textbf{PCGCN}~\cite{pcgcn};
\circledw{4} GNN frameworks and accelerators: \textbf{DGL}~\cite{dgl}, \textbf{PyG}~\cite{pyg}, and \textbf{GNNAdvisor}~\cite{wang2021gnnadvisor}. 
We use the latest open-source implementations of these methods as strong baselines.
Table~\ref{tab:precision} summarizes precision and computation resources supported by these works.

\textbf{Datasets: } 
We use the same set of 500 representative large sparse matrices as the FlashSparse test set (all from SuiteSparse~\cite{suitesparse}).
In addition, we also select classic graph datasets across different application domains such as IGB~\cite{khatua2023igb}, Reddit~\cite{hamilton2017inductive} for end-to-end GNN performance and preprocessing overhead (as shown in Table~\ref{tab:gnndataset}).

\definecolor{mygreen}{HTML}{009900} 
\newcommand{\cmark}{\color{mygreen}\faCheck}
\newcommand{\xmark}{\color{red}\faTimes}
\begin{table}[]
    \centering
    \caption{Capability comparison of Baselines and Libra}
    \begin{tabular}{cccccc}
    \toprule
    \multirow{2}[2]{*}{Works} &  \multicolumn{3}{c}{Precision} &   \multicolumn{2}{c}{Computation}  \\ 
     \cmidrule(lr){2-4}     \cmidrule(lr){5-6}
      & FP32 & TF32 & FP16 & TCUs & CUDA cores \\
    \midrule
    TC-GNN &\xmark & \cmark & \xmark &  \cmark & \xmark\\
    DTC-SpMM & \xmark & \cmark &  \xmark & \cmark & \xmark\\
    FlashSparse &\xmark & \cmark & \cmark & \cmark & \xmark\\
    cuSPARSE  & \cmark & \xmark & \xmark & \xmark & \cmark\\
    Sputnik  & \cmark & \xmark & \xmark  & \xmark & \cmark\\
    RoDe  & \cmark & \xmark &  \xmark & \xmark & \cmark\\
    DGL & \cmark & \xmark &  \xmark & \xmark & \cmark\\
    PyG  & \cmark & \xmark &  \xmark & \xmark & \cmark\\
    GNNAdvisor & \cmark & \xmark &  \xmark & \xmark & \cmark\\
    HC-SpMM  & \cmark & \cmark & \xmark  & \cmark & \cmark\\
    RT-GNN  & \cmark & \cmark & \xmark  & \cmark & \cmark\\
    PCGCN & \cmark & \cmark & \xmark  & \cmark & \cmark\\
    {\bf Libra} & \cmark & \cmark & \cmark  &  \cmark & \cmark\\
    \bottomrule
    \end{tabular}
    \label{tab:precision}
\end{table}

\begin{table}[!ht]
    \centering
    \caption{Datasets for GNNs evaluation}
    \begin{tabular}{cccc}
    \toprule
    Dataset & \#Vertex &  \#Edge &  \#AvgRowL\\
    \midrule
    IGB-small & 1,000,000 & 13,068,130 & 13.07 \\ 
    Reddit & 232,965 & 114,848,857 & 492.9\\ 
    Amazon & 403,394 & 9,068,096 & 22.48\\ 
    \bottomrule
    \end{tabular}
        \label{tab:gnndataset}
\end{table}


\subsection{SpMM and SDDMM Kernel Performance}
\begin{figure}[htbp]
  \centering
  \includegraphics[width=\columnwidth]{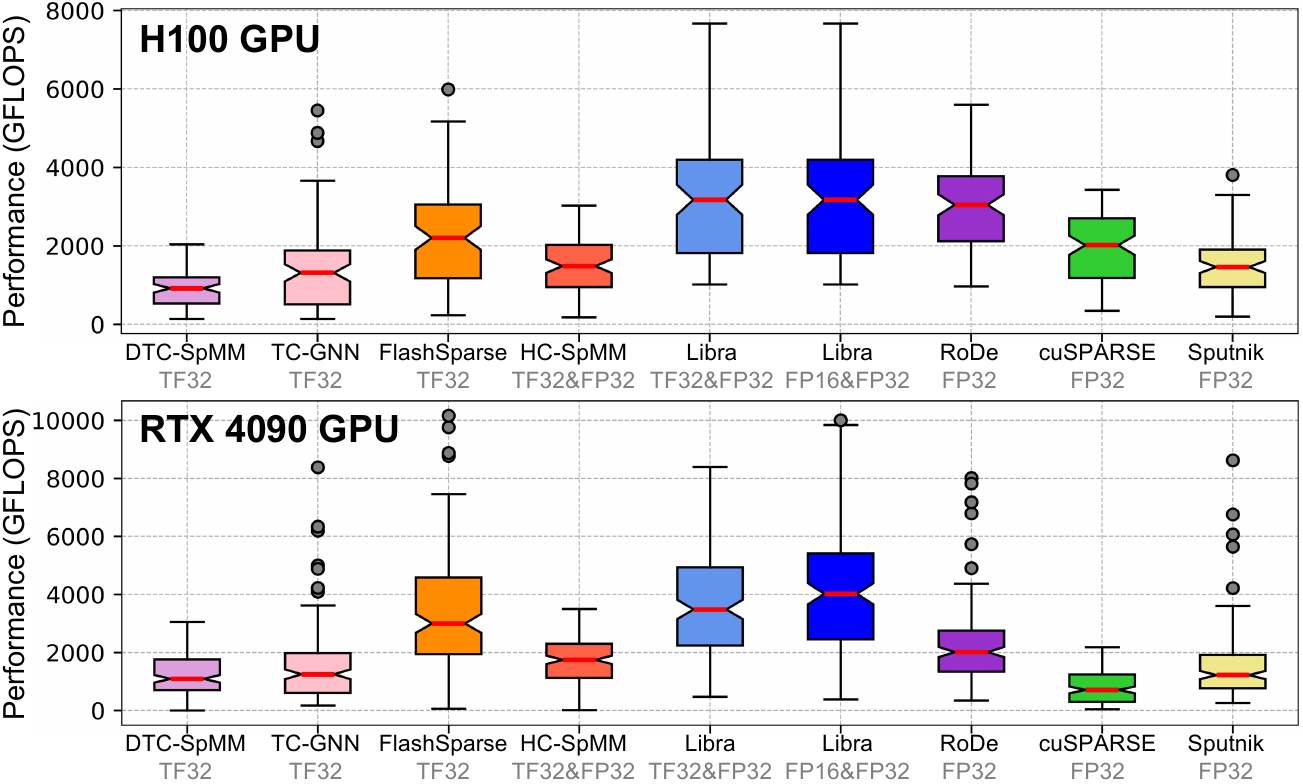}
    \caption{SpMM throughput on H100 and RTX 4090 GPUs.}
    \label{fig:spmm}
\end{figure}

Figure~\ref{fig:spmm} presents the throughput distribution of SpMM across different GPU architectures, measured over 500 sparse matrices, with N set to a commonly used value of 128 (i.e., the number of columns in the dense matrix B).
As illustrated in Figure~\ref{fig:spmm}, Libra outperforms all baselines in both precisions across the majority of matrices on the H100 and RTX 4090 GPUs.
As the number of nonzero elements grows, Libra’s performance advantage becomes more pronounced.
We further summarize detailed SpMM speedup results of Libra (TF32\&FP32) over competitive baselines (using TF32 or FP32) in Table~\ref{tab:spmm}. The experimental results show that Libra achieves geometric mean speedups of 1.73$\times$ (up to 4.69$\times$) over RoDe (SOTA on CUDA cores) on the RTX 4090 GPU, 2.94$\times$ (up to 10.65$\times$) over DTC-SpMM, and 1.77$\times$ (up to 8.63$\times$) over FlashSparse (SOTA on TCUs) on the H100 GPU.
These performance improvements result from Libra’s hybrid computation, which achieves improved TCU utilization (evaluated in Section~\ref{sec:ab_1}) and enhanced data reuse (evaluated in Section~\ref{sec:ab_2}).
Moreover, compared to the state-of-the-art hybrid approach HC-SpMM, Libra achieves a geometric mean speedup of 5.29$\times$ (up to 50$\times$). 
This performance improvement primarily results from our precise 2D-aware workload distribution (validated in Section~\ref{sec:eff_2D}) and decoupled execution paths for TCUs and CUDA cores, enabling flexible task scheduling and customized kernel implementations (validated in Sections~\ref{sec:ab_4} and~\ref{sec:ab_5}).

\setlength{\tabcolsep}{5pt}
\begin{table*}[htbp]
    \centering
    \caption{Speedup distribution of SpMM for Libra (TF32\&FP32) over Baselines (TF32 or FP32)}
\begin{tabular}{ccccccccccccc}
\toprule
     \multirow{2}[2]{*}{\textbf{Baselines}} & \multicolumn{6}{c}{\textbf{H100}} & \multicolumn{6}{c}{\textbf{RTX 4090}} \\
     \cmidrule(lr){2-7} \cmidrule(lr){8-13} 
     & $\boldsymbol{<}\textbf{1x}$ & \textbf{1\textasciitilde1.5x} & \textbf{1.5\textasciitilde2x} & $\boldsymbol{\geq}\textbf{2x}$ & \textbf{Mean} & \textbf{Max} & $\boldsymbol{<}\textbf{1x}$ & \textbf{1\textasciitilde1.5x} & \textbf{1.5\textasciitilde2x} & $\boldsymbol{\geq}\textbf{2x}$ &  \textbf{Mean} & \textbf{Max}\\
\midrule
cuSPARSE & 1.05\% & 31.86\% & 39.87\% & 27.22\% & 1.81x & 28.48x &  0.21\% & 5.03\% &11.32\% & 83.44\% & 7.06x & 75.05x \\
RoDe & 43.46\% & 37.13\% & 16.46\% &2.95\% & 1.13x & 3.37x & 6.71\% & 27.67\% & 30.40\% &35.22\% & 1.73x & 4.69x \\
DTC-SpMM & 0.3\% & 2.95\% & 14.14\% & 82.61\% & 2.94x & 10.65x & 1.03\% & 4.73\% & 8.85\% & 85.39\% & 3.12x & 9.16x \\
FlashSparse & 0.21\% & 68.9\% & 9.49\% & 21.4\% & 1.77x & 8.63x & 19.75\% & 62.75\% & 3.09\% & 14.41\% & 1.34x & 5.06x \\
HC-SpMM & 1.13\% & 41.69\% & 34.37\% & 22.81\% & 5.29x & $\geq50$x & 7.22\% & 48.34\% & 18.61\% & 25.83\% & 4.10x & $\geq50$x \\

\bottomrule
    \end{tabular}
    \label{tab:spmm}
\end{table*}

\begin{figure}[htbp]
  \centering
  \includegraphics[width=\columnwidth]{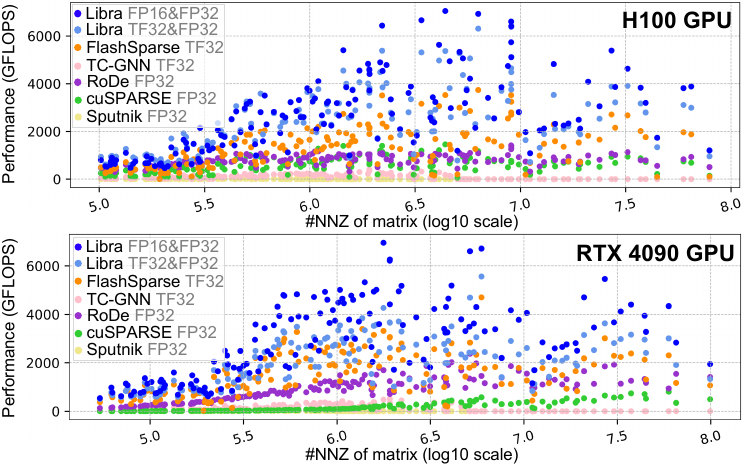}
    \caption{SDDMM throughput on H100 and RTX 4090 GPUs. Each point shows the mean GFLOPS of three matrices.}
    \label{fig:sddmm}
\end{figure}

Moreover, since TC-GNN and Sputnik exhibit low SDDMM throughput that is unsuitable for box plots, we use scatter plots instead (Figure~\ref{fig:sddmm}, N=32, where N denotes columns in dense matrices A and B). Libra consistently outperforms baselines. Notably, TC-GNN and Sputnik perform poorly on matrices with over 5 million nonzero elements, and their GFLOPS are labeled as 0.
The experimental results show that Libra (TF32\&FP32) achieves geometric mean speedups of 2.73$\times$ (up to 21.09$\times$) over FlashSparse on H100 GPU, and 3.05$\times$ (up to 8.98$\times$) over RoDe on RTX 4090 GPU.

\subsection{Effectiveness of 2D-Aware Workload Distribution}
\label{sec:eff_2D}

\subsubsection{Comparison of workload distribution strategies.}
\begin{figure}[htbp]
  \centering
  \includegraphics[width=0.95\columnwidth]{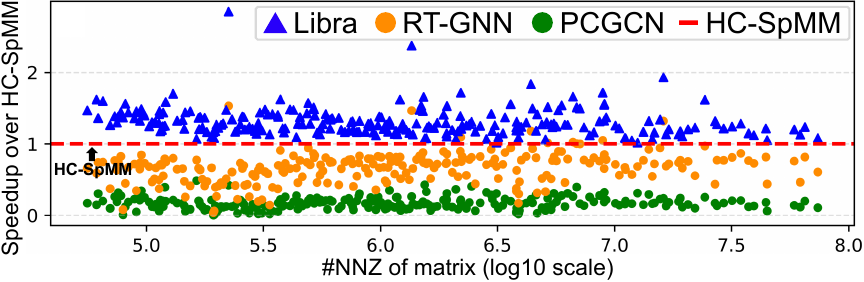}
    \caption{ Libra SpMM performance with different workload distribution strategies (all using TF32\&FP32 precision).}
    \label{fig:compare_strategies}
\end{figure}

To isolate the effect of workload distribution strategies, we evaluate Libra against baselines that employ identical kernel implementations from Libra but differ solely in their workload distributions (HC-SpMM, RT-GNN, and PCGCN) across 500 matrices.
As shown in Figure~\ref{fig:compare_strategies}, Libra achieves average speedups of 1.31$\times$ (up to 4.34$\times$) over HC-SpMM, 3.22$\times$ (up to 47.03$\times$) over RT-GNN, and 10.1$\times$ (up to 502$\times$) over PCGCN.
These improvements stem from our strategy integrating data reuse (theoretical) and TCU utilization (empirical) for precise workload distribution across heterogeneous resources.


\subsubsection{Comparison with single-resource computation.}
\label{sec:compare_single}
To clarify in which cases Libra performs well or poorly relative to single-resource execution, we evaluate three execution modes (CUDA-core-only, TCU-only, and hybrid) across 500 matrices, identifying the best-performing mode for each.
For SpMM, the hybrid approach outperforms single-resource modes on 369 matrices, with average speedups of $1.59\times$ over CUDA-core-only and $1.22\times$ over TCU-only. For SDDMM, hybrid execution achieves the best performance on 453 matrices, yielding average speedups of $1.97\times$ over CUDA-core-only and $1.19\times$ over TCU-only.
These results indicate that Libra’s hybrid computation excels for matrices combining structured and unstructured sparsity, but underperforms for exclusively structured or unstructured matrices.

\subsubsection{The selection of threshold $\eta_{thr}$.}
\label{sec:threshold_eva}
\begin{figure}[htbp]
  \centering
  \includegraphics[width=\columnwidth]{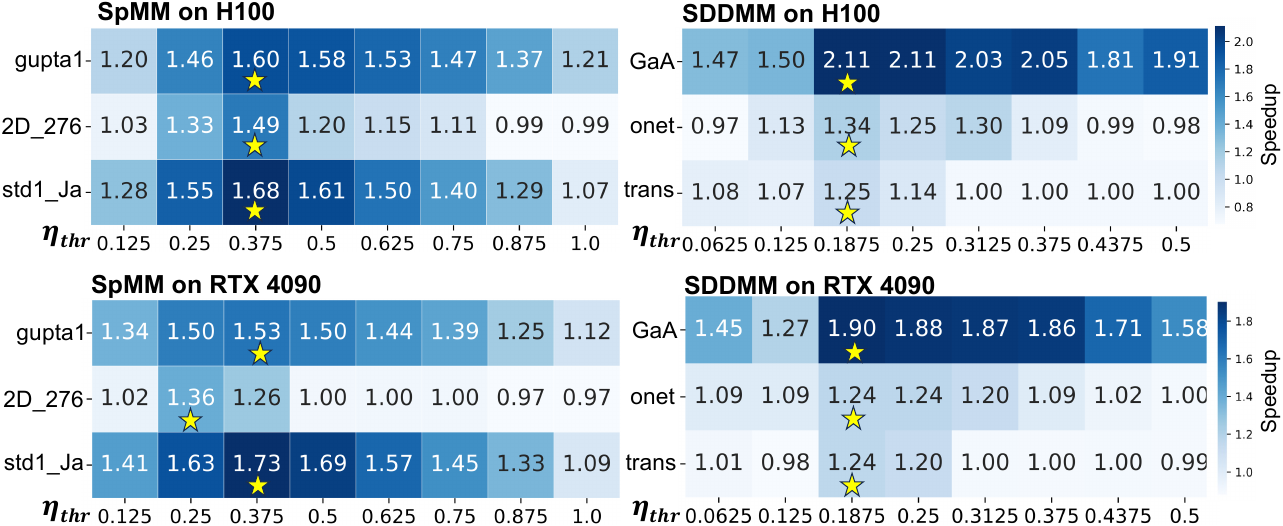}
  \caption{The optimal threshold to achieve the best performance across different operators and different matrices. The speedup is compared to the CUDA-core-only pattern.}
  \label{fig:threshold}
\end{figure}

The $\eta_{thr}$, defined in Section~\ref{sec:practice}, denotes the minimum TCU utilization required to assign workloads to TCUs.
For SpMM, the threshold $\eta_{thr}$ ranges from 0.125 to 1.0 for an 8 $\times$ 1 vector, while for SDDMM, we evaluated eight commonly used threshold $\eta_{thr}$ values from 0.0625 to 0.5 in increments of 0.0625.
As shown in Figure~\ref{fig:threshold}, we select representative matrices exhibiting significant hybrid acceleration in Section~\ref{sec:compare_single}. The optimal thresholds $\eta_{thr}$, marked by stars, are 0.375 for SpMM and 0.1875 for SDDMM, providing maximum speedups.
This indicates that similar threshold selections are effective across different matrices, empirically validating our theoretical analysis in Section~\ref{sec:practice}. 
For a given architecture, the threshold $\eta_{thr}$ can be determined once and reused across different matrices.
Consequently, compared with existing approaches~\cite{pcgcn,ye2023sparsetir, li2025hc} that require extensive manual tuning with numerous parameters (e.g., PCGCN relying on METIS tuning, HC-SpMM using a logistic-regression model), our 2D-aware workload distribution strategy precisely identifies optimal task mapping with negligible tuning overhead.

\setlength{\tabcolsep}{10pt}
\begin{table*}[htbp]
    \centering
    \caption{Speedup distribution in ablation studies; \textbf{\#Effective} denotes the number of matrices that outperform the baselines}
    \begin{tabular}{ccccccc}
    \toprule
    \multirow{2}[2]{*}{\textbf{Components}} &  \multirow{2}[2]{*}{\textbf{\#Effective}} &  
      \multicolumn{2}{c}{\textbf{Speedup}}  &   \multirow{2}[2]{*}{\textbf{Mean}} &  \multirow{2}[2]{*}{\textbf{Max}}\\ 
     \cmidrule(lr){3-4}
      & & \textbf{1x-1.2x} &  $\boldsymbol{\geq}\textbf{1.2x}$ & & \\
    \midrule
    Load balancing in SpMM & 467/500 &  67.4\%  &32.6\%  & 3.8x & 102x\\
    Load balancing in SDDMM & 359/500 &  5.3\%  &94.7\%  & 29x & 155x\\
    Multi-stream scheduling mode & 291/500 &  77.7\% &22.3\% & 1.5x & 1.8x\\
    Sequential scheduling mode & 209/500 &  85.2\% &14.8\% & 1.1x & 1.8x\\
    Shared memory bypassing in SpMM & 500/500 & 71.2\% & 28.8\%  & 1.2x & 1.3x\\
    Shared memory bypassing in SDDMM & 309/500 & 95.7\% & 4.3\%  & 1.1x & 1.3x\\
    Bitmap in SDDMM & 490/500 & 1.8\% & 98.2\%  & 2.6x & 21x \\
    Preprocessing (GPU vs. OpenMP) & 491/500 &  0.4\%  &99.6\%  & 17x &  395x\\
    \bottomrule
    \end{tabular}
    \label{tab:ab_format}
\end{table*}

\subsection{Ablation Studies}
\label{sec:ab}
In this section, we analyze 500 matrices on the H100 GPU to identify key factors behind Libra’s performance gains.
\subsubsection{Improvement of TCU utilization.}
\label{sec:ab_1}
\begin{figure}[htbp]
  \centering
  \includegraphics[width=\columnwidth]{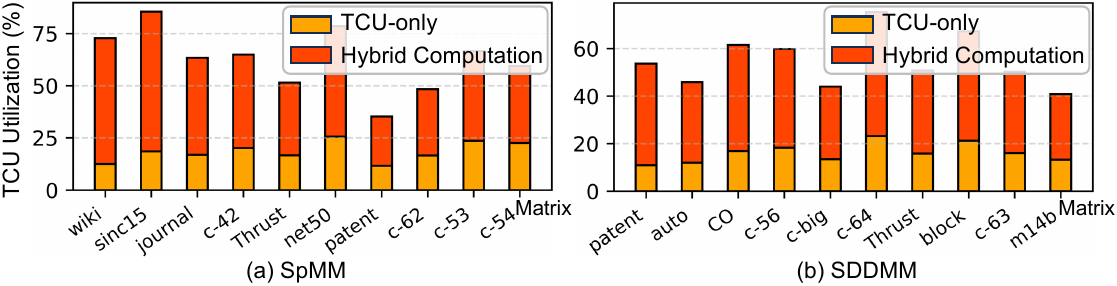}
    \caption{Comparison of TCU utilization under Hybrid and TCU-only execution modes in Libra.}
    \label{fig:ab-tcu}
\end{figure}

We compare TCU utilization under hybrid and TCU-only execution modes. Hybrid computation achieves an average TCU utilization improvement of $1.39\times$ (up to $5.8\times$) for SpMM, and $1.56\times$ (up to $4.90\times$) for SDDMM. Figure~\ref{fig:ab-tcu} shows the comparison for representative matrices. These results confirm that hybrid computation significantly enhances TCU utilization by distributing extremely sparse workloads to CUDA cores, thereby increasing TC block density.

\subsubsection{Data access reduction.}
\label{sec:ab_2}
\begin{figure}[htbp]
  \centering
  \includegraphics[width=\columnwidth]{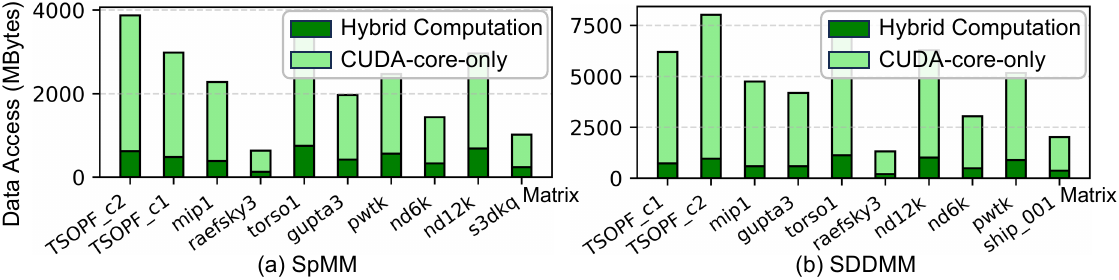}
    \caption{Comparison of Data Access Cost under Hybrid and CUDA-core-only execution modes in Libra.}
    \label{fig:ab-data}
\end{figure}

We compare data access costs (fetching operands from L1/L2/global-memory) between hybrid and CUDA-core-only execution modes.
For SpMM, hybrid computation reduces data access by an average of 49.73\% (up to 86.19\%). For SDDMM, the average reduction is 54.19\% (up to 89.45\%).
Representative matrices are illustrated in Figure~\ref{fig:ab-data}.
These results confirm that hybrid computation significantly reduces data access by distributing denser workloads to TCUs, thus exploiting the data reuse capabilities of MMA operations.

\subsubsection{The effectiveness of load balancing.}
\label{sec:ab_3}
As shown in Table~\ref{tab:ab_format}, the hybrid load balancing strategy yields notable average speedups. Moreover, it improves SM utilization by an average of $2.3\times$ compared to the unbalanced baseline. These results demonstrate that our strategy effectively balances heterogeneous workloads across thread blocks, significantly improving concurrency, despite the atomic-operation overhead from workload decomposition.

\subsubsection{The effectiveness of task scheduling.}
\label{sec:ab_4}
Table~\ref{tab:ab_format} compares the two scheduling modes, indicating that multi-stream scheduling performs better on 291 matrices, whereas sequential scheduling is preferred on 209 matrices. Our occupancy-aware strategy dynamically selects between these modes, ensuring optimal hybrid execution efficiency.

\subsubsection{The effectiveness of Bitmap.}
\label{sec:ab_5}
We exploit temporal locality when loading bit-encoded elements to fully bypass shared memory. As shown in Table~\ref{tab:ab_format}, compared to caching bit-encoded elements in shared memory with intra-block synchronization~\cite{fan2025spinfer,accspmm,xia2025voltrix}, our approach improves performance for both SpMM and SDDMM. 
Moreover, for SDDMM, our Bitmap-based write-back approach achieves an average $2.6\times$ speedup over the iterative-traversal-based ME-TCF~\cite{dtc}.

\subsection{Case Study: End-to-End GNN Performance}
\begin{figure}[htbp]
  \centering
  \includegraphics[width=0.95\columnwidth]{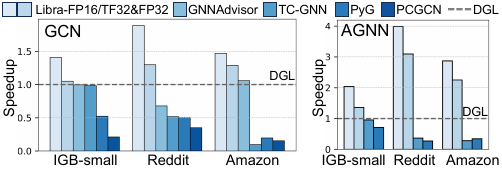}
    \caption{End-to-end GNN performance on H100 GPU.}
    \label{fig:gnn}
\end{figure}

To evaluate end-to-end performance, we select two mainstream GNN models: GCN~\cite{kipf2016semi} and AGNN~\cite{thekumparampil2018attention}, each configured with five layers and trained for 300 epochs.
As shown in Figure~\ref{fig:gnn}, Libra outperforms the baseline (DGL) on both GCN and AGNN, achieving geometric mean speedups of $1.57\times$ (up to $1.89\times$) for GCN and $2.9\times$ (up to $3.9\times$) for AGNN.

\subsection{Preprocessing Overhead}
Moreover, the preprocessing in Libra include 2D-aware workload distribution, load balancing, and format translation.
To demonstrate the efficiency of our GPU-accelerated preprocessing, we further compare it with an OpenMP-based CPU implementation. As shown in Table~\ref{tab:ab_format}, the average speedup is $17.1\times$\xspace, with a maximum speedup of $395.8\times$\xspace.
Notably, the preprocessing overhead becomes even more negligible during end-to-end GNN inference and training.
For instance, in the GCN evaluation, preprocessing accounts for only 0.4\% of the total training time.
As the number of layers and epochs increases, the relative cost of preprocessing decreases further.
This is because preprocessing only needs to be performed once, and the results can be reused in subsequent tasks.


\section{Related Work}
Numerous efforts have been made to accelerate SpMM and SDDMM operators across GNNs~\cite{wang2023tc,dtc,flashsparse,chen2025accelerating,chen2025groot,accspmm}, scientific computing~\cite{okanovic2024high,koanantakool2016communication,liu2024unisparse,hong2019adaptive,RoDe}, and large language models (LLMs)~\cite{xia2023flash,fan2025spinfer,wang2025generalsparse}. 
With the introduction of TCUs in NVIDIA's Volta architecture, researchers began leveraging their powerful matrix computation capabilities, initially targeting structured sparsity scenarios~\cite{magicube,chen2021efficient}.
The rising popularity of GNNs, characterized by inherently sparse and unstructured data, has driven recent efforts to efficiently map irregular sparse computations onto TCUs.
Works such as TC-GNN~\cite{wang2023tc}, DTC-SpMM~\cite{dtc}, and FlashSparse~\cite{flashsparse} redesigned data formats and introduced specialized tiling techniques, enabling efficient mapping of irregular sparse workloads onto TCUs. 
Nevertheless, even highly optimized TCU-only methods still suffer from computational redundancy under extremely sparse workloads.
In contrast, CUDA cores offer a more flexible programming model, making them better suited for efficiently handling irregular sparse workloads~\cite{huang2020ge,cuSPARSE,RoDe}.
This motivates hybrid solutions, such as PCGCN~\cite{pcgcn}, RT-GNN~\cite{yan2025rt}, and HC-SpMM~\cite{li2025hc}, which aim to leverage both resources effectively.
However, existing hybrid methods lack a systematic analysis of hybrid computation gains and a clear definition of complementarity between TCUs and CUDA cores, hindering theoretically and empirically driven workload distribution and efficient task mapping.
To close these gaps, we propose a fine-grained, 2D-aware workload distribution method along with efficient task scheduling and kernel implementations.



\section{Conclusion}
Libra brings new insights for accelerating sparse workloads by utilizing the on-chip heterogeneous computing resources on GPUs. Especially, based on the deep analysis to identify the real advantages of hybrid computation, we propose a novel 2D-aware (locality and utilization) workload distribution method to precisely select the optimal task mapping for different sparse workloads (SpMM and SDDMM). Libra also integrates the techniques of hybrid load balancing and occupancy-aware task scheduling for high-performance kernel implementation. Our approach and technical scheme enable Libra to achieve the best of both worlds, namely combining the high data reuse of TCUs with the low computational redundancy of CUDA cores. Extensive experiments on H100 and RTX 4090 GPUs demonstrate that Libra achieves state-of-the-art performance for sparse kernels and end-to-end applications. Although our work is conducted on NVIDIA GPUs, the approach of Libra is also applicable to other GPU architectures (such as AMD GPUs, DCUs, and NPUs) with heterogeneous computing resources. We foresee that our approach will form a new paradigm for accelerating sparse workloads with heterogeneous computing resources on GPUs.

\section{Acknowledgement}
This project was supported by the National Natural Science Foundation of China under Grant No. 62372055, the National Science and Technology Major Project (2023ZD0120502), and the Fundamental Research Funds for the Central Universities.

\bibliographystyle{ACM-Reference-Format}
\bibliography{sample-base}

\end{document}